\documentclass[preprint,showpacs,floats,letterpaper,floatfix,,groupedaddress,eqsecnum]{revtex4}

\usepackage{subfigure}

\usepackage{amssymb,amsmath}
\usepackage[dvips]{graphicx}
\usepackage{csquotes}

\usepackage{dcolumn,epsfig}
\usepackage[left]{lineno}

\begin{document}

\title{Resonance oscillation of a damped driven simple pendulum}

\author{D. Kharkongor$^1$$^,$$^2$, and Mangal C. Mahato$^1$$^,$}
\email{mangal@nehu.ac.in}
\affiliation{$^1$Department of Physics, North-Eastern Hill University,
Shillong-793022, India}
\affiliation{$^2$Department of Physics, St. Anthony's College,
Shillong-793003, India}

\begin{abstract}
The resonance characteristics of a driven damped harmonic oscillator are 
well known. Unlike harmonic oscillators which are guided by parabolic 
potentials, a simple pendulum oscillates under sinusoidal potentials. The 
problem of an undamped pendulum has been investigated to a great extent. 
However, the resonance characteristics of a driven damped pendulum have not 
been reported so far due to the difficulty in solving the problem analytically.
In the present work we report the resonance characteristics of a driven damped 
pendulum calculated numerically. The results are compared with the resonance 
characteristics of a damped driven harmonic oscillator. The work can be of 
pedagogic interest too as it reveals the richness of driven damped motion of a 
simple pendulum in comparison to and how strikingly it differs from the motion 
of a driven damped harmonic oscillator. We confine our work only to the 
nonchaotic regime of pendulum motion.
\end{abstract}


\maketitle
\section{Introduction}

The motion of a simple pendulum has the same equation as the motion of a 
particle in a sinusoidal potential. And the motion of the simple pendulum with 
very small amplitude $x$ (so that $\sin x$ can be safely approximated as 
$x$) is equivalent to that of a harmonic oscillator. However, as the amplitude
becomes larger, the motion of the simple pendulum differs from that of a harmonic 
oscillator. The natural frequency of oscillation of a harmonic oscillator 
($\omega_0=\sqrt{\frac{k}{m}}$) is independent of its amplitude, with $m$ as 
the mass of the oscillator and $k$ is the stiffness constant of the spring. 
However, since a simple pendulum with large amplitude is different from a 
harmonic oscillator its frequency of oscillation is not independent of the 
amplitude of its free oscillation. The period or the frequency, $\omega$, of 
oscillation of a freely oscillating pendulum of finite amplitude, $x_0$, is 
given in terms of an elliptic integral of the first kind and has been given in 
many text books, as for example\cite{Sommerfeld,Kittel,Marion}. The frequencies
in terms of simple series expansions have also been given\cite{Kittel,Marion} 
and there are many attempts to improve upon the series expansion in order 
to get to as close as the exact (elliptic integral) value using only few terms,
to cite a few\cite{Kidd,Lima,Rajesh,Johannessen,Belendez,Salas}. For comparison
of the frequency of oscillation of harmonic oscillator, $\omega_0$, and various
expresssions for the frequency, $\omega_1$, of oscillations of a simple 
pendulum we use the expressions used by Kittel, et. al.\cite{Kittel}
\begin{equation}
\frac{\omega_1}{\omega_0}\approx 1-\frac{x_0^2}{16}
\end{equation}
and the expression arrived at by Bel\'{e}ndez, et. al.\cite{Belendez} 
\begin{equation}
\frac{\omega_1}{\omega_0}\approx 
\frac{1}{4}\left(1+\sqrt{\cos\frac{x_0}{2}}\right)^2
\end{equation}
in addition to the exact result\cite{Belendez},
\begin{equation}
\frac{\omega_1}{\omega_0}=\frac{\pi}{2K(k)},
\end{equation}
where
\begin{equation}
K(k)=\int_0^{\frac{\pi}{2}}\frac{d\phi}{\sqrt{1-k\sin^2\phi}}
\end{equation}
and
\begin{equation}
k=\sin^2\frac{x_0}{2}
\end{equation}
and plotted in Fig. 1, using standard tables\cite{Abramowitz} for the elliptic 
integrals (1.4). It is to be noted that whereas the frequency of free 
oscillation of a simple harmonic oscillator is independent of amplitude, the 
frequency of free oscillation of a simple pendulum decreases with its amplitude.

It is also well known that if a harmonic oscillator is driven by an external 
periodic drive of frequency $\omega$, then the oscillator responds with larger
and larger amplitude $x_0$ as $\omega\rightarrow\omega_0$ and the amplitude
becomes infinitely large at $\omega=\omega_0$. The condition $\omega=\omega_0$
is termed as the resonance condition. Note, however, that a simple pendulum 
cannot have a similar resonance oscillation at a fixed frequency 
$\omega=\omega_0$ as $\omega(x_0)$ is not independent of $x_0$. 
 
When the harmonic oscillator is (viscous) damped so that it satisfies the 
equation of motion\cite{Marion,Kleppner} 
\begin{equation}
\frac{d^2x}{dt^2}+\gamma\frac{dx}{dt}+\omega_0^2x=0,
\end{equation}
the oscillator, for small damping coefficient $\gamma < 2\omega_0$, oscillates 
with displacement
\begin{equation}
x(t)=Ae^{\frac{-\gamma t}{2}}\cos(\omega_1t+\phi),
\end{equation}
with the frequency of oscillation 
$\omega_1=\sqrt{\omega_0^2-\frac{\gamma^2}{2}}$ 
and diminishing amplitude $Ae^{-\frac{\gamma t}{2}}$, where $A$ and $\phi$ are
arbitrary constants. Notice again that even though frequency of oscillation 
depends on the damping factor $\gamma$, it remains fixed for all time, 
independent of the (diminishing) amplitude. The motion of (under)damped 
pendulum can be described by 
\begin{equation}
\frac{d^2x}{dt^2}+\gamma\frac{dx}{dt}+\omega_0^2\sin x=0.
\end{equation}
An approximate solution has been derived by Johannessen\cite{Johannessen1}. The
approximate solution compares quite well with the numerical solution. The
solution is oscillatory in nature with diminishing amplitude as time 
progresses, Fig. 1 of Ref.\cite{Johannessen1}. The time periods of oscillations
are plotted in Fig. 3 of the same refrence as a function of time. One can 
clearly notice that initially the frequency of oscillations are small but keep
increasing with time and the frequency approaches the value appropriate for the 
damped harmonic oscillator, $\omega_1=\sqrt{\omega_0^2-\frac{\gamma^2}{2}}$.
This is as it should be because as the amplitude of oscillation approaches zero
the motion of simple pendulum becomes closer to that of a simple harmonic 
oscillator. 

The above description reiterates the well-known result that, in presence of 
damping, the harmonic oscillator as well as the simple pendulum asymptotically 
approach the stationary state $x(t\rightarrow \infty )=0$. However, if the 
damped harmonic oscillator is, in addition, subjected to a periodic forcing 
$F(t)=F_0\cos \omega t$, then the equation of motion 
\begin{equation}
\frac{d^2x}{dt^2}+\gamma\frac{dx}{dt}+\omega_0^2x=\frac{F_0}{m}\cos \omega t
\end{equation}
has a solution 
\begin{equation}
x(t)=x_0\cos (\omega t+\phi),
\end{equation}
with
\begin{equation}
x_0=\frac{F_0}{m}\frac{1}{((\omega^2-\omega_0^2)^2+
(\omega\gamma)^2)^{\frac{1}{2}}}
\end{equation}
and
\begin{equation}
\phi=\tan^{-1}(\frac{\gamma\omega}{\omega^2-\omega_0^2}),
\end{equation}
is the phase difference between $F(t)$ and the response $x(t)$. Owing to the
presence of damping a mean power loss of
\begin{equation}
P=\frac{F_0^2}{2m}\frac{\gamma\omega}{(\omega^2-\omega_0^2)^2+(\omega\gamma)^2}
\end{equation}
occurs. $P$ shows a peak exactly at the frequency $\omega=\omega_0$, even 
though the amplitude peaks at a lower frequency 
$\omega=\sqrt{\omega_0^2-\frac{\gamma^2}{2}}$. $P$ becomes maximum at 
$\omega=\omega_0$ because the phase difference is $\frac{\pi}{2}$ at this
frequency. One can take this frequency $\omega=\omega_0$ as the resonance 
frequency of the forced damped harmonic oscillator. Obviously, the resonance
frequency is independent of the damping coefficient or the amplitude of
oscillation. The corresponding resonance condition for a damped simple pendulum
driven by a periodic force does not have a known analytical expression. We,
therefore, numerically obtain the resonance frequency of the forced damped 
simple pendulum as discussed in the following.

\section{Resonance Frequency}

We investigate the behavior of a forced damped simple pendulum in the present 
work. In particular, we explore the resonance behavior of such a pendulum by 
calculating the corresponding mean power loss or equivalently the hysteresis 
loss. We, in fact, investigate the equivalent problem of motion of an 
underdamped particle in a sinusoidal potential under the influence of an 
external periodic force $F(t)=F_0\cos \omega t$.

The equation of motion in a periodic potential $V(x)=-\sin x$ is given by (in
dimensionless form)
\begin{equation}
\frac{d^2x}{dt^2}+\gamma\frac{dx}{dt}-\cos x=F_0\cos \omega t.
\end{equation}
Note that this is exactly the equation of a damped driven simple pendulum if
one identifies the angular displacement $\theta=x-\frac{\pi}{2}$. By solving
this equation we obtain the mean hysteresis loop, its area, and the amplitudes
of the corresponding trajectories $x(t)$ in order to characterize the resonance
behavior of the system. We identify, as explained earlier, the resonance
frequency as one for which the hysteresis loop area is the largest and 
calculate the corresponding mean amplitude $\overline{x_0}$ of oscillation
$x(t)$. The mean hysteresis loop area is just the integral $\int F dx$ over a
large number $N$ of periods of $F(t)$ and calculate the average for one cycle
of $F$. This quantity is again averaged over $M$ initial conditions $x(t=0)$
but with $v(t=0)=\frac{dx}{dt}(t=0)=0$, giving the mean hysteresis loop area
$\overline{A}$, for various values of frequency $\omega$ for a given $\gamma$. 
We set $F_0=0.2$ throughout in this work except in the inset of Fig. 2. 
We keep the dimensionless mean amplitude $F_0$ small so that the pendulum always remain in the nonchaotic regime.
We, thus, identify the frequency $\overline{\omega_0}$ which gives the largest 
hysteresis loop area. The corresponding mean amplitude of oscillation 
$\overline{x_0}$ is also calculated. Fig. 2 gives the plot of the resonance 
frequency $\overline{\omega_0}$ versus the mean amplitude $\overline{x_0}$. 
Each point on the plot corresponds to a fixed value of $\gamma$. For 
comparison, we also plot the resonance frequency $\omega=\omega_0$ of 
oscillation of damped driven harmonic oscillator in the same graph. Also 
plotted in the same graph is the exact frequency of free ($\gamma=0$) 
oscillation of the simple pendulum as a guide to the eye. A similar graph is 
plotted for drive amplitude $F_0=0.1$ as an inset to Fig. 2.

\begin{figure}[htp]
\centering
\includegraphics[width=15cm,height=11cm]{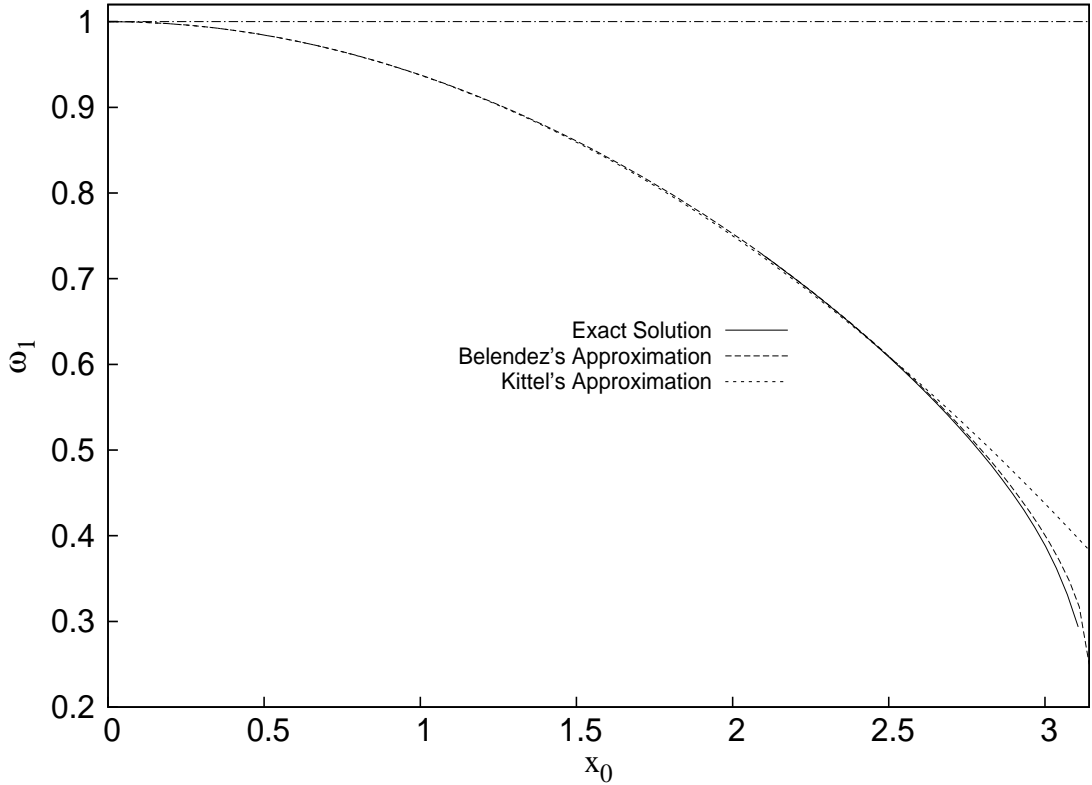}
\caption{The figure shows the frequency of oscillation $\omega_1$ as
a function of initial position $x_0$ for an undamped simple pendulum.
As is shown in the figure, notice that the approximations as obtained by
Kittel and Bel\'{e}ndez depart from the exact result at large amplitudes.
Also, shown in the figure (by the dash-dotted horizontal line)
is the corresponding frequency of oscillation $\omega_0$
of an driven-damped harmonic oscillator as a function of initial position $x_0$.}
\end{figure}

In Fig. 2 at least three qualitatively different regions can be clearly identified. 
One, in the intermediate range $0.165\leq\gamma\leq 0.38$ of $\gamma$, the 
obtained graph shows the usual tendency of increasing resonance frequency as the 
amplitude of oscillation decreases, as though to approach the harmonic 
oscillator limit, as $\gamma$ increases. And the tendency is more so for the 
smaller drive amplitude $F_0=0.1$ as shown in the inset of the figure. Of 
course, the range of $\gamma$ in this case does not lie in 
$0.165\leq\gamma\leq 0.38$ but $0.099\leq\gamma\leq 0.28$. Two, for large 
$\gamma>0.38$, the obtained $\overline{\omega_0}$ decreases rapidly with 
decreasing amplitude $\overline{x_0}$ as $\gamma$ increases. 
$\overline{\omega_0}$ thus moves away from the harmonic oscillator limit of
resonance frequency $\omega_0$. In both these regions of $\gamma$ the resonance
frequencies remain lower than the free oscillation frequencies. And thirdly, in
the small $\gamma$ regime, the resonance frequency abruptly drops to a small 
value just below $\gamma \approx 0.165$, thus we get a disjoint branch of the 
resonance amplitude frequency curve. Thereafter $\overline{\omega_0}$ decreases much slowly with increasing $\overline{x_0}$. The slow decrease of 
$\overline{\omega_0}$ allows the curve to ultimately cross the 
amplitude-frequency curve of free oscillation of the simple pendulum. Thus, at smaller
$\gamma$ values the resonance frequency $\overline{\omega_0}$ becomes larger
than the free oscillation frequency. In the following, we discuss the resonance
oscillations in the three regions of $\gamma$ separately.

\begin{figure}[htp]
\centering
\includegraphics[width=15cm,height=11cm]{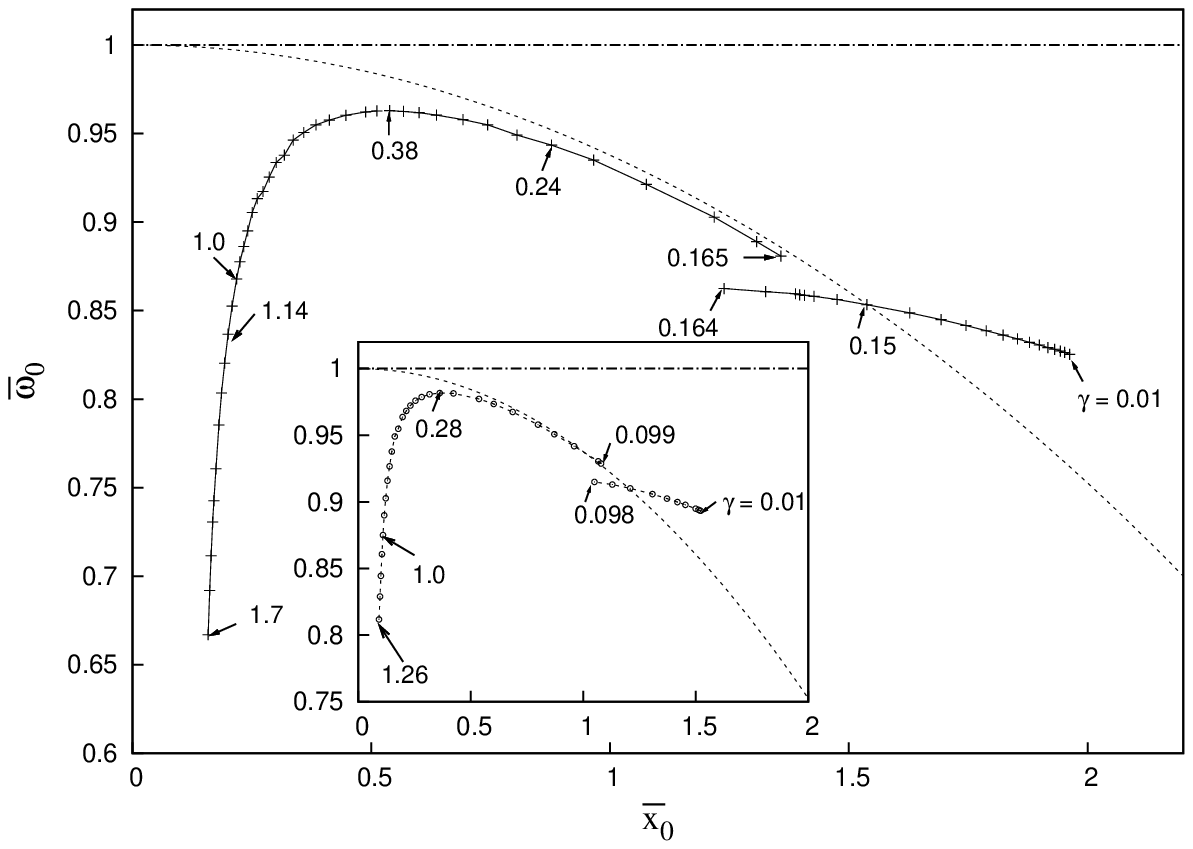}
\caption{The figure shows the plot of the resonance
frequency $\overline{\omega_0}$ versus the mean amplitude $\overline{x_0}$
of the damped-driven pendulum with each point corresponding to a particular
value of damping $\gamma$ (a few of them labelled) for a forcing amplitude $F_0 = 0.2$.
The inset shows the same but with a forcing amplitude $F_0 = 0.1$. Also shown,
in both the main plot as well as in the inset, are the exact result of the
frequency of oscillation $\omega_1$ of an undamped simple pendulum (by broken line)
and (by the dash-dotted horizontal line) the corresponding resonance frequency of oscillation $\omega_0$
of an driven-damped harmonic oscillator.}
\end{figure}

\begin{figure}[htp]
\centering
\includegraphics[width=15cm,height=11cm]{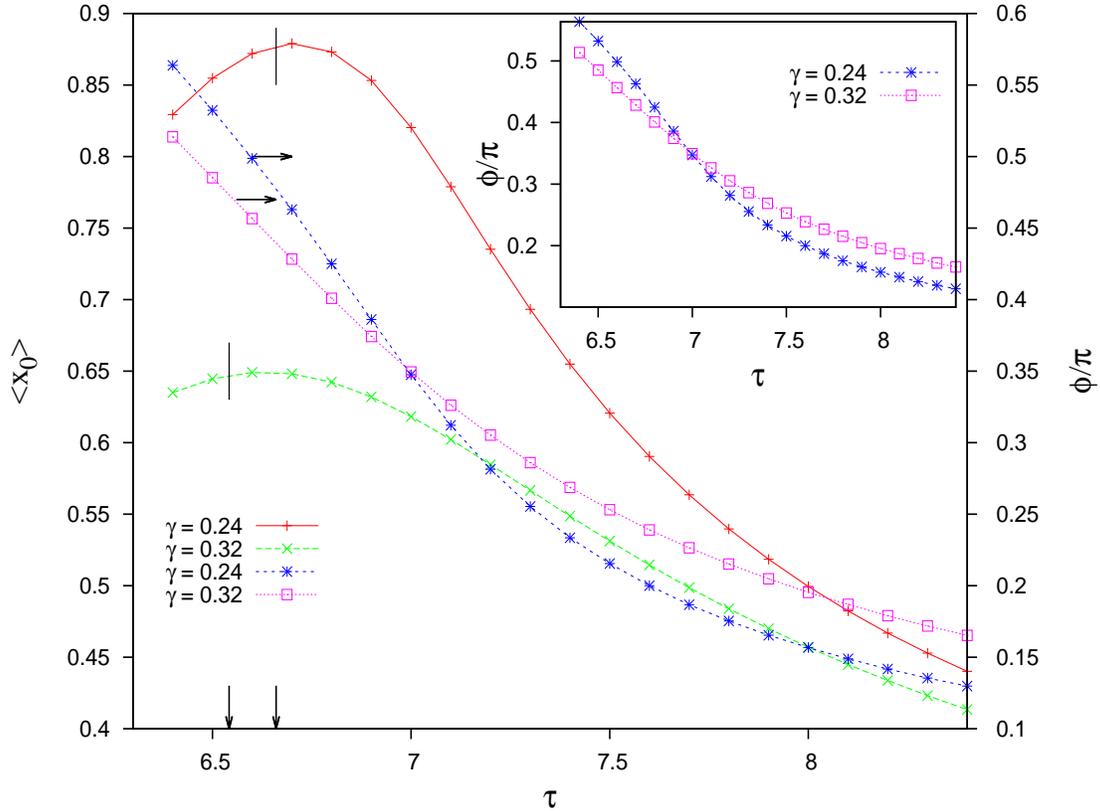}
\caption{The figure shows the variation of the amplitude of the ensemble
averaged hysteresis loop and the phase of the loop as a function of the driving
period $\tau$. Note that the two graphs indicated by the horizontal arrows
(y-axis label is towards the right hand side) signify the phase of the loop
for the two $\gamma$ values as indicated. The phase variation as
a function of driving period is replotted in the inset.
The right vertical arrow and the right
vertical line on $<{x_0}>$ indicates the resonant period, $\tau_0$ = $6.66$ for $\gamma = 0.24$ whereas the left
vertical arrow and the left vertical line on $<{x_0}>$ is for $\gamma = 0.32$ with  $\tau_0$ = $6.542$.}
\end{figure}

\begin{figure}[htp]
\centering
\includegraphics[width=15cm,height=11cm]{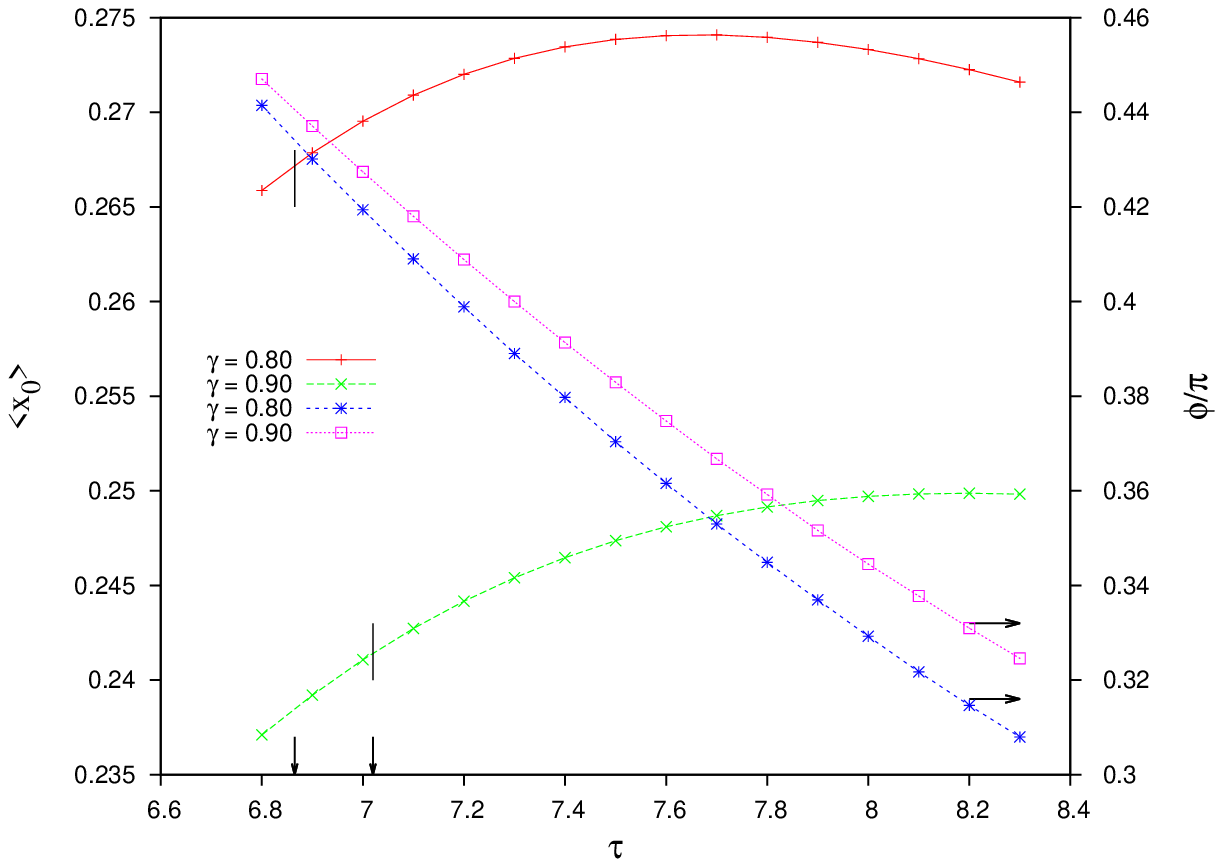}
\caption{The figure shows the variation of the amplitude of the ensemble
averaged hysteresis loop and the phase of the loop as a function of the
driving period $\tau$. Note that the two graphs indicated by the horizontal
arrows (y-axis label is towards the right hand side) signify the phase of the
loop for the two $\gamma$ values as indicated. The right vertical arrow
and the right vertical line on $<{x_0}>$
indicates the resonant period, $\tau_0$ = $7.02$ for $\gamma = 0.90$ whereas
the left vertical arrow and the left vertical line on $<{x_0}>$
is for $\gamma = 0.80$ with  $\tau_0$ = $6.865$.}
\end{figure}

Figure 2 is the main result of our paper. It shows the contrasting resonance
behavior of a linear and a nonlinear system. Usual physical intuition often
does not work in nonlinear systems. It is, therefore, hard to explain offhand
why a simple pendulum behaves so differently from a harmonic oscillator as far
as the resonance characteristics are concerned. We only try to provide a rough 
explanation using our numerical results.

The hysteresis loop area $\int Fdx$ depends on both the amplitude $<x_0>$ of the 
response trajectory $x(t)$ as well as the phase difference $\phi$ between 
$F(t)$ and $x(t)$. For intermediate as well as large damping $\gamma$, the 
phase difference $\phi$ decreases with increasing period $\tau$ of $F(t)$ for all
$\gamma$ values. Also, the amplitude $<x_0>$ is smaller for larger $\gamma$ 
values for any period $\tau$ of the drive. These are understandable. However, 
as we can see from Figs. 3 and 4, the amplitude $<x_0>$ shows a nonmonotonic 
behavior, it peaks at some intermediate $\tau$ value. This is a peculiar 
feature in the present case of periodically driven underdamped sinusoidal 
(nonlinear) potential system.

\subsection{Intermediate range of $\gamma$ values}

This range refers to $0.165\leq\gamma\leq0.38$, of $\gamma$ for $F_0=0.2$ and 
$0.099\leq\gamma<0.28$ for $F_0=0.1$. Both present similar features though we
consider Fig. 3 for the former case only. Figure 3 shows both the amplitude
$<x_0>$ and the phase lag $\phi$ together, for the two typical values of $\gamma$ equal 
to 0.24 and 0.32, for easy comparison. However, the inset of the figure has 
only the plot of phase lag as a function of the drive period $\tau$. The 
peculiar features to be noticed are that $<x_0>$ shows a peak in the lower $\tau$
range for both the $\gamma$ values. However, the peak corresponding to 
$\gamma=0.32$ occurs at a slightly lower value of $\tau$ than $\gamma=0.24$. 
Moreover, in the same small range of $\tau$ the phase lag $\phi$ is larger in 
the case of $\gamma=0.24$ than $\gamma=0.32$. On the other hand, in the larger 
range of $\tau$ the phase lag $\phi$ is larger for the larger $\gamma=0.32$. 
Thus in the larger range of $\tau$ the phase lag shows the usual behaviour. 
The small values of $\phi$ in this range shows that when the system is driven 
at a slow rate the system $x(t)$ follows the drive $F(t)$ closely. However, the
same does not hold in the range of smaller $\tau$ (or larger frequency 
$\omega$) where inertia seems to play an important role.

In the larger frequency $\omega$ range it is difficult for the system $x(t)$ to
follow the drive $F(t)$ and consequently the phase lag $\phi$ between them is
large, $\phi$ close to $\frac{\pi}{2}$. Moreover, $x(t)$ seems to overshoot
$F(t)$. That is, while $F(t)$ turns back from its maximum $x(t)$ continues on
its way for a longer while, before it retraces its path back due to its inertia.
And inertia is more effective at lower damping. However, at still larger 
frequencies the system again fails to respond to the field and the amplitude 
begins to decrease. With this plausible qualitative explanation of the peaking 
behaviour of $<x_0>$ as a function of $\tau$ it becomes easier to see why the 
resonance frequency $\overline{\omega_0}$ increases with decreasing amplitude
$\overline{x_0}$.   

As mentioned earlier, $\overline{\omega_0}$ corresponds to the drive frequency
$\omega$ at which the mean power loss is maximum and $\overline{x_0}$ is the
corresponding amplitude of $x(t)$. And, $F(t)$ being sinusoidal and $x(t)$ 
also being roughly sinusoidal, given an amplitude $<x_0>$ of $x(t)$ the power 
loss or the hysteresis $F(x)$ loop area becomes maximum when 
$\phi=\frac{\pi}{2}$. If $\phi$ is nearly $\frac{\pi}{2}$ it is the amplitude 
$<x_0>$ that determines the maximum of the hysteresis loss. Therefore in this 
case maximum of power loss occurs at a frequency close to where $<x_0>$ becomes 
maximum. From Fig. 3, one can, therefore, see that $\overline{\omega_0}$ is 
smaller for $\gamma=0.24$ than $\gamma=0.32$ with correspondingly larger 
$\overline{x_0}$ for $\gamma=0.24$ than $\gamma=0.32$. Thus, considering these 
two typical values of $\gamma$, in this intermediate range of $\gamma$, the 
curve $\overline{\omega_0}(\overline{x_0})$ should have the same qualitative
nature as given in Fig. 2, that is, $\overline{\omega_0}$ decreases with 
increasing $\overline{x_0}$. However, the variation is not so sharp as in the 
large $\gamma$ case ($\gamma>0.38$ for $F_0=0.2$).

\subsection{Large $\gamma$ regime}

In this large damping ($\gamma>0.38$ for $F_0=0.2$ and $\gamma>0.28$ for 
$F_0=0.1$) regime naturally amplitudes $<x_0>$ of oscillation are relatively 
small compared to those in the smaller $\gamma$ regimes. It is also intuitively
obvious that the response amplitude should decrease with increasing damping, 
for example, $<x_0>(\gamma=0.8)$ is greater than $<x_0>(\gamma=0.9)$. That the resonance frequency 
should also decrease with damping can be qualitatively explained again from 
the obtained variation of $<x_0>$ and $\phi$ with $\tau$ as in Fig. 4. 

We choose two large values of $\gamma$ equal to 0.8 and 0.9 for illustration. 
Here, the phase lags $\phi$ are small ($\phi<\frac{\pi}{2}$) and $\phi$ is 
consistently smaller for $\gamma=0.8$ than $\gamma=0.9$ at any value of $\tau$.
The effect of inertia gradually diminishes as $\gamma$ increases. The 
consequence of this can be seen from the diminishing sharpness of $<x_0>$ peaks 
with increasing $\gamma$. Also to be noticed from Fig. 4 is that in case of 
$\gamma=0.8$ the $<x_0>$ peak occurs at larger frequency (smaller $\tau$) than 
$\gamma=0.9$. The variation of $<x_0>$ is much smaller than the variation of 
$\phi$ in the plotted relevant region of $\tau$. The resonance peak occurs 
close to where the hysteresis loop area is the largest, that is where $\phi$ 
is large ($\approx \frac{\pi}{2}$), thus from the figure, close to small values
of $\tau$ (but larger than in case of intermediate range of $\gamma$ considered 
earlier). Both $<x_0>$ and $\phi$ conspire together to maximize the hysteresis 
loop area to determine $\overline{\omega_0}$ and corresponding $\overline{x_0}$
at resonance. Fig. 4, thus helps in getting a plausible qualitative idea that 
at larger damping not only the resonance response amplitudes is smaller but 
the system becomes slower to respond too. The rapid rise of 
$\overline{\omega_0}$ with $\overline{x_0}$ with decreasing $\gamma$ is an 
outcome that does not seem unusual as $\gamma$ decreases $\overline{\omega_0}$ 
tends towards the natural frequency of 
oscillation $\omega_1$ at $\gamma=0$, Fig. 2.

\subsection{Small $\gamma$ regime}

For small $\gamma$ values, for example $0.06\leq\gamma < 0.165$ for $F_0=0.2$, 
the explanation of the nature of the resonance curve 
$\overline{\omega_0}(\overline{x_0})$ has an entirely different origin. The
resonance curve is disjoint from the earlier two regimes (in which the curves
were contiguous). In this range of $\gamma$, the hysteresis loop area maximizes 
as a function of frequency in a region of frequency where there exists not one 
kind of particle trajectory but two for a given $\gamma$ and amplitude $F_0$ of
$F(t)$ \cite{SaikiaX,WandaX,Decay}. Of course, if the amplitude $F_0$ is large
(say, $>0.25$), the trajectories become chaotic\cite{Wanda1X}. We consider only 
amplitudes $F_0$ that give nonchaotic trajectories.

\begin{figure}[htp]
\centering
\includegraphics[width=15cm,height=12cm]{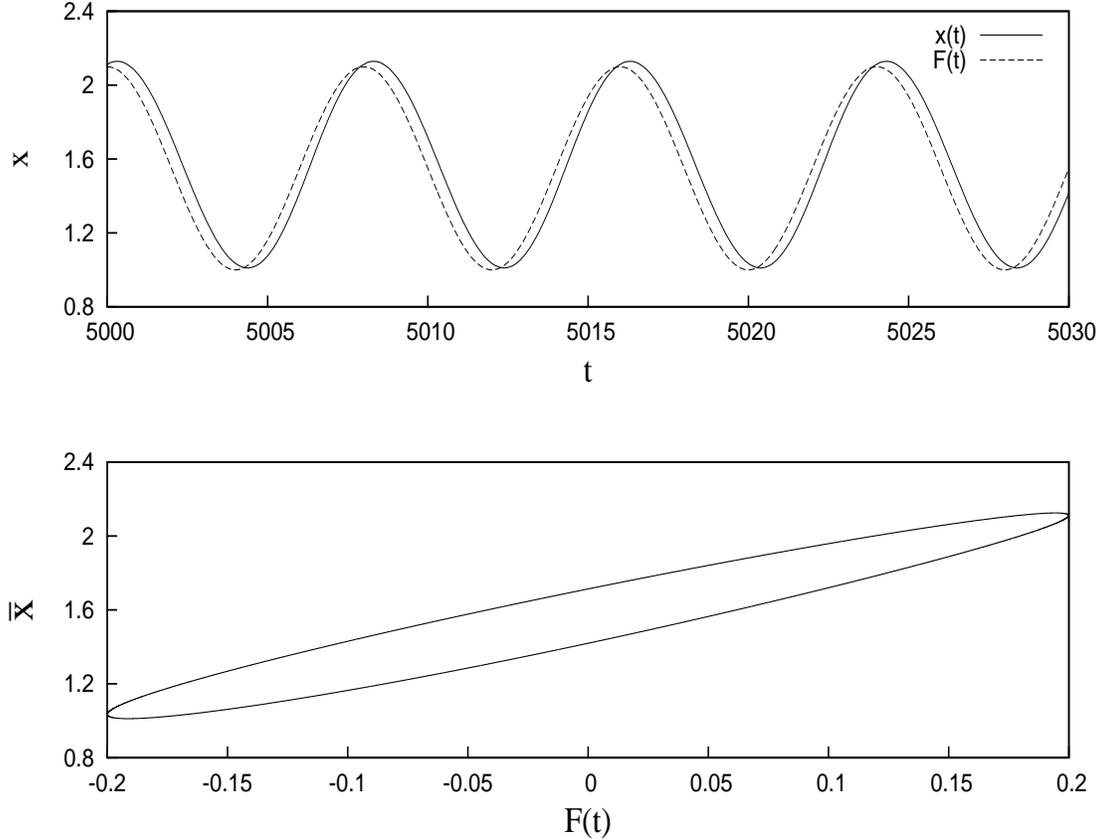}
\caption{The top panel represents an SA state with a small phase lag with 
respect to the forcing $F(t)$ with amplitude $F_0 = 0.2$. Notice that the 
amplitude $F_0$ in the top panel is magnified so that comparison with $x(t)$ 
can be visualised easily. The bottom panel represents the corresponding 
hysteresis loop.}
\end{figure}

\begin{figure}[htp]
\centering
\includegraphics[width=15cm,height=12cm]{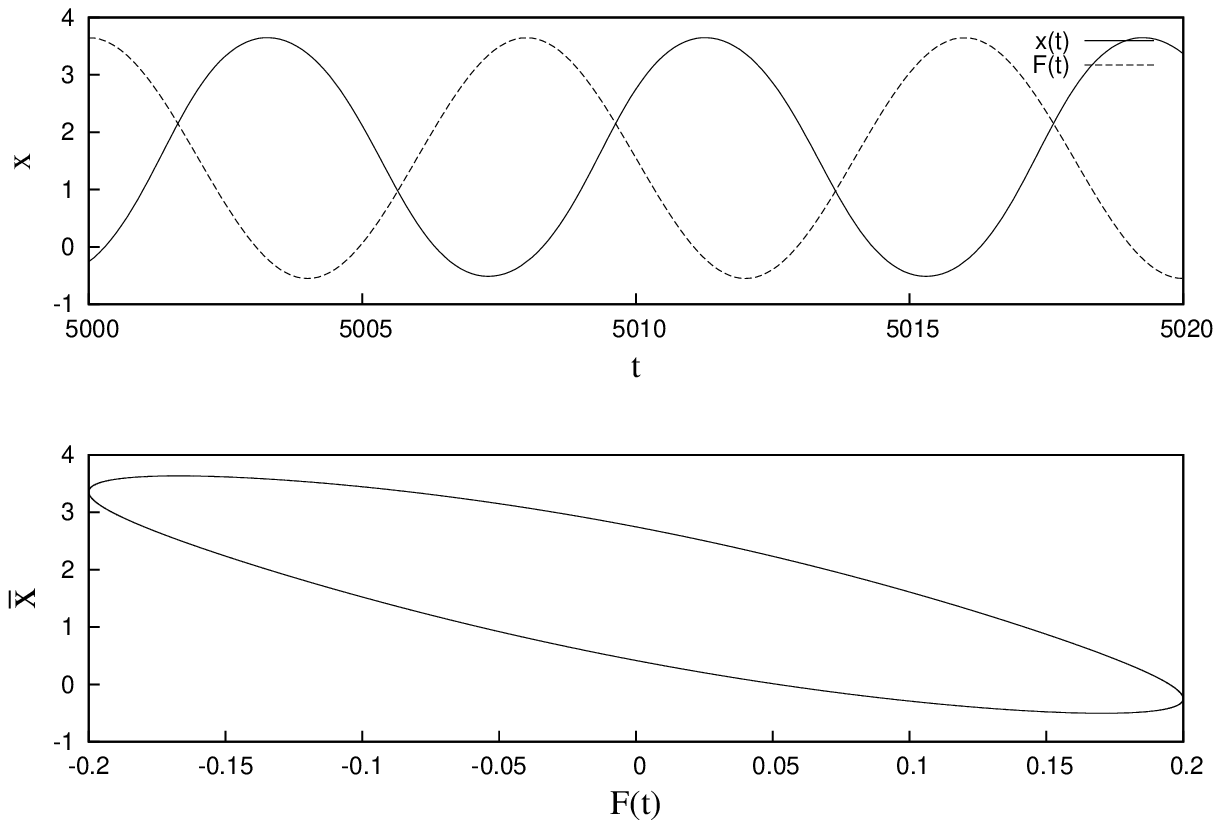}
\caption{The top panel represents an LA state with a larger phase lag with 
respect to the forcing $F(t)$ with amplitude $F_0 = 0.2$. Notice that the 
amplitude $F_0$ in the top panel is magnified so that comparison with $x(t)$ 
can be visualised easily. The bottom panel represents the corresponding 
hysteresis loop.}
\end{figure}

\begin{figure}[htp]
\centering
\includegraphics[width=12cm,height=15cm,angle=270]{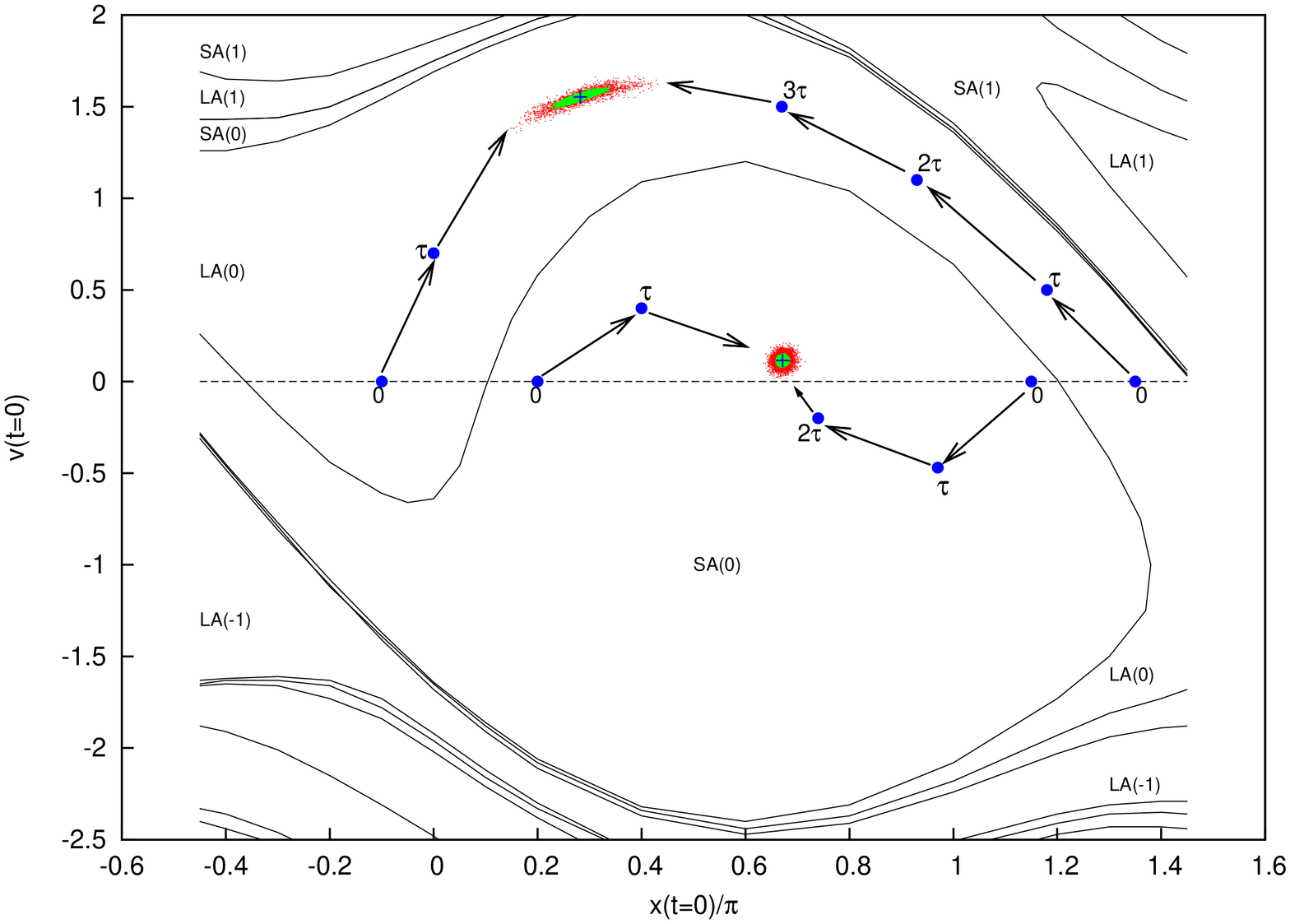}
\caption{The basins of attraction of the LA and SA attractor in the absence 
(cross-mark) and presence of minute fluctuations is shown. The bracketed 
numbers on LA and SA indicate the well number of the periodic potential where 
the trajectory settles down in the LA attractor or the SA attractor. For this 
figure, $\gamma = 0.12$, $\tau = 8.0$ and $F_0 = 0.2$ and $V(x)=-\sin x$. The 
horizontal line in the middle is when the particles initially have velocity 
$v(t=0) = 0$.  Corresponding to this zero initial velocity, the LA attractors 
are the upper colored regions whereas the SA attractors are the lower colored 
regions. The green regions correspond to $T = 0.0001$ and the red regions 
correspond to $T = 0.001$.}
\end{figure}

These two states of trajectory have the status of distinct dynamical states 
having well defined basins of attraction in the ($x(0),v(0)$) space. One of 
the two states of trajectories has a comparatively small amplitude (SA) and 
has a small phase lag ($\phi<\frac{\pi}{3}$) with respect to the external drive 
$F(t)$ whereas the other state has a large amplitude (LA) and a large phase 
lag. Figures 5 and 6 show the SA and LA states with their respective mean
hysteresis loops shown in the bottom panels of both the figures. 
To obtain the hysteresis loop, as explained earlier, the trajectory of the pendulum is averaged over the entire 
duration of the trajectory as $\overline{x}(F(t))$.
The area bounded by the hysteresis loops corresponds to the 
energy dissipated to the surrounding medium. The basins of attraction of these 
two dynamical states with parameter values $\gamma = 0.12$, $\tau = 8.0$ and 
$F_0 = 0.2$ are shown earlier in \cite{SaikiaX}. We reproduce an updated plot 
for explanation purposes in Fig. 7. In Fig. 7 we also show the stroboscopic 
plots in the $(x(0)-v(0))$ plane when the initial velocity of the particles at 
$t=0$ is $v(t=0) = 0$ and in the absence and presence of minute thermal 
fluctuations. These stroboscopic plots appear in Fig. 7 as colored regions. In 
the absence of fluctuations, these stroboscopic regions reduce to a single 
point, corresponding to an attractor, as shown by the cross-mark. The upper 
stroboscopic plot corresponds to the LA attractor while the lower stroboscopic 
plot corresponds to the SA attractor. The basins of attraction shown in Fig. 7 
have the usual physical significance. For example, if the initial 
position of the particle were to lie in the range of, say, 
$0.2\pi\le x(0)\le 1.1\pi$, then on evolving the system after initial transients
have died out, the particle will home into the SA attractor and will remain in 
its initial well only. But on the other hand, suppose if the initial position 
of the particle were to lie in the range of, say, $-0.3\pi \le x(0) \le 0$, then
the particle will oscillate with larger amplitude and get fixed to the LA 
attractor. The (transient) time evolutions towards the fixed centres of 
attractions have been shown schematically in Fig. 7 and marked by arrows. We 
note that the illustration given is when the initial velocity of the particle 
at any position within one period of the potential well is zero. This zero 
initial velocity is represented in Fig. 7 as a horizontal line.

\begin{figure}[htp]
\centering
\includegraphics[width=15cm,height=11cm]{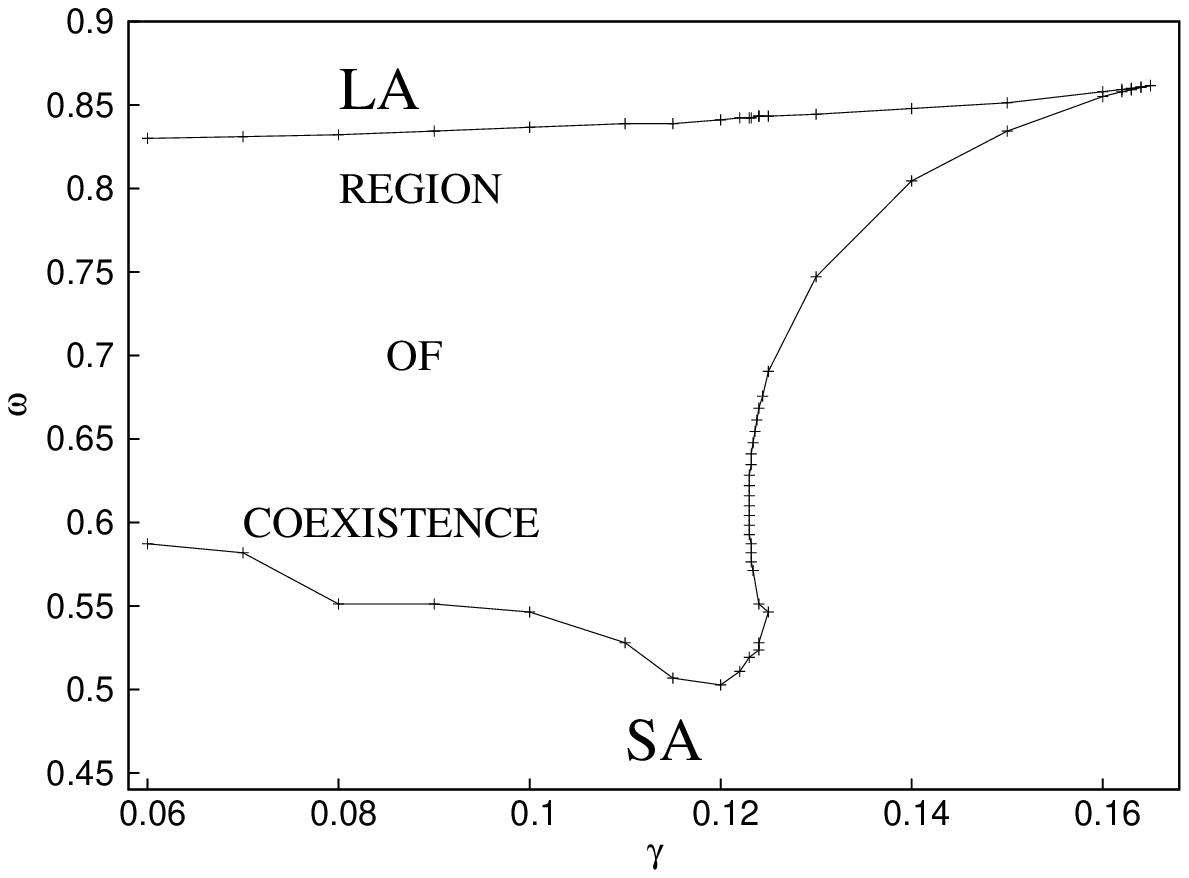}
\caption{Phase diagram in the $\omega$-$\gamma$ space denoting the regions of 
pure LA, pure SA, and coexistence region. Here, $F_0 = 0.2$.}
\end{figure}

\begin{figure}[htp]
\centering
\includegraphics[width=15cm,height=11cm]{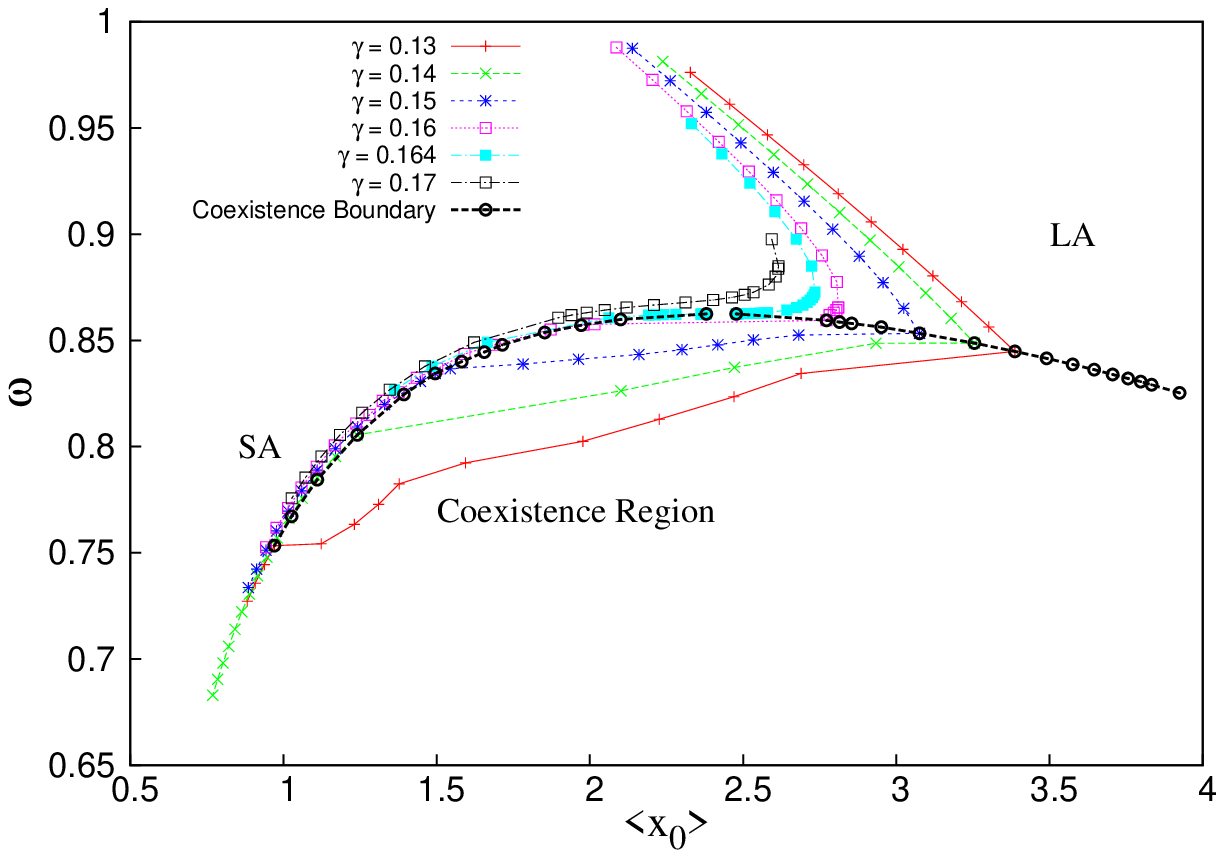}
\caption{The figure shows the variation of the amplitudes of the LA state and 
the SA state as a function of the driving frequency $\omega$. The boundary 
line (dash-open circle line) separate the coexistence region from the 
regions where either SA state is present or LA state is present only. The 
amplitudes between the boundary line are calculated in the coexistence region 
for $\gamma = 0.13$, $\gamma = 0.14$ and $\gamma = 0.15$.}
\end{figure}

For a given $\gamma$ the domain of the basins of attraction depends on the 
value of $\omega$, Fig. 8. For small $\omega$ (for example, for 
$\gamma=0.13,~\omega<0.74$) only SA states appear whereas for large $\omega$ 
(for the same $\gamma,~\omega>0.85$) the domain of the basins of attraction of 
SA states shrink to zero and only LA states appear and for the intermediate 
$\omega$ both states coexist. We choose initial velocity $v(0)=0$ always and 
hence the occurrence of LA or SA states depends on the initial position 
($x(0)$) within a period of the sinusoidal potential. In the region of 
coexistence of the two states we choose two hundred initial positions lying in 
the range $\frac{-\pi}{2}<x(0)\leq\frac{3\pi}{2}$ at equal intervals in order 
to calculate the mean values that takes into account the presence of the two 
dynamical states in right proportions. At this point it should be noted that 
the hysteresis loop area corresponding to the LA states are larger compared to 
the area corresponding to the SA states and hence the need to consider all 
possible values of $x(0)$ in order to get the mean values of the hysteresis 
loop area and the mean amplitude $<x_0>$ of the trajectories.

For $\gamma>0.165$, in the intermediate range of $\gamma$ for $F_0=0.2$, there 
is only one kind of trajectory for all values of $\omega$ and initial positions
$x(0)$, the mean amplitude $<{x_0}>$ varies continuously with the 
varation of $\omega$. It turns out that $\gamma\approx 0.164$ is the critical 
value of $\gamma$ above which there is no distinction between the LA and SA 
states. However, for $\gamma<0.164$, as mentioned earlier, for large value of 
$\omega$ we obtain only LA states but as $\omega$ is decreased some LA states 
corresponding to some initial positions $x(0)$ give way to SA states. The 
particular value of $\omega$, depending on the value of $\gamma$, at which SA 
states begin appearing also gives the largest hysteresis loop area and can be 
identified with $\overline{\omega_0}$ and the corresponding $<x_0>$ as the 
$\overline{x_0}$. The locus of these ($\overline{x_0},\overline{\omega_0}$) is 
shown in Fig. 9 by the right-hand side dash-open circle thick boundary line. This boundary line 
is the resonance line shown in Fig. 2 for $\gamma<0.165$. At 
$\gamma\approx 0.164$, $\overline{\omega_0}$ abruptly drops to a smaller value 
due to the sudden appearance of the SA states of trajectories (in place of some 
earlier LA trajectories) with mean hysteresis loop area of SA states being much 
smaller than LA states. The slopes of ($\overline{x_0},
\overline{\omega_0}$) lines of Fig. 9 abruptly change indicating the beginning 
of the coexistence region of the two states.

\subsection{A superficial analogy}

Fig. 9 is quite instructive. Apart from the thick boundary line on the large
$<x_0>$ side of the abscissa, we have an another thick
boundary line on the small $<x_0>$ side. For smaller $<x_0>$ 
values we obtain only SA states of the pendulum. These two boundary
lines, one separating the LA states from the coexistence region and the other
separating the coexistence region from the purely SA region, meet at a point 
where the slope of the $(<x_0>,\omega)$ line is zero for $\gamma\approx 0.164$. 
This point is somewhat analogous to the critical point in the ($P-\rho$) 
diagram of the liquid-gas system. The two thick boundary lines thus enclose 
the region of coexistence of the two states and  separate the LA states from 
the SA states of trajectories. We show in Fig. 9, for $\gamma=0.13$, 
$\gamma=0.14$ and $\gamma = 0.15$, the mean amplitude of the trajectories when the 
particles are in the pure LA state, pure SA state and when the states coexists.
For calculating the amplitude in the coexistence region, we ensemble average 
over all initial conditions taken and obtain the ensemble averaged hysteresis 
loop whereby the ensemble averaged amplitude can be calculated.

Though analogy of the SA and LA states of trajectories of the pendulum with
 the liquid-gas phase is quite superficial it is suggestively 
tempting to draw further analogy between the system of two dynamical states, 
LA and SA, of trajectories with the phases of the liquid-gas system. The 
frequency $\omega$ appears analogous to the pressure $P$, the friction 
coefficient $\gamma$ to the temperature $T$ and the amplitude $<x_0>$ of the 
trajectory to the density $\rho$ of the liquid. And finally the hysteresis 
loop area $\overline{A}(\omega,\gamma)$ seems analogous to the Gibbs free 
energy $g(P,T)$ of the liquid.
\begin{equation}
\begin{array}{c}
\omega\longleftrightarrow P\\
\gamma\longleftrightarrow T\\
<x_0>\longleftrightarrow\rho\\
\overline{A}(\omega,\gamma)\longleftrightarrow g(P,T)
\end{array}
\end{equation}  

However, these analogies are not exact but only superficial and cannot be 
stretched far.

\section{Concluding remarks}

In this work we have investigated the dynamics of a driven damped simple 
pendulum or equivalently the motion of a particle in a sinusoidal potential in 
a medium that offers dissipation. Note that we have investigated the motion of 
the simple pendulum at the temperature $T=0$, that is, without considering the 
effect of thermal fluctuations. Also, we have kept the drive amplitude $F_0$
small so that the motion is nonchaotic.

In order to investigate the motion of a damped driven simple pendulum one needs
to go much beyond the usual LCR circuit problem. In fact, no analytical 
solution has so far been found for this problem. The numerical solutions 
obtained, however, offer interesting insight. In the underdamped regime, in 
certain range of $\gamma$ and amplitude and frequency of the drive $F(t)$, two 
distinct solutions exist for the same periodic driving force $F(t)$. This 
leads to an entirely new relationship between the amplitude of motion with the 
resonance frequency of the simple pendulum having no relationship with the 
corresponding driven damped simple harmonic oscillator. That is, the resonance 
characteristics of the damped harmonic oscillator cannot be extrapolated to 
obtain the resonance characteristics of a damped simple pendulum. 

\section*{Acknowledgement}
 We thank the Computer Centre, North-Eastern Hill University, Shillong, for 
providing the high performance computing facility, SULEKOR.

\end{document}